 \documentclass[final,5p,times,twocolumn,authoryear]{elsarticle}

\usepackage{amssymb}
\usepackage{lipsum}
\usepackage{xcolor}
\usepackage{comment}
\usepackage[T1]{fontenc}

\journal{TBA}

\begin{document}

\begin{frontmatter}



\title{
A Note on the Quality of Dilatonic Ultralight Dark Matter
}
\author[b]{Jay Hubisz}
\author[c]{Shaked Ironi}
\author[c]{Gilad Perez}
\author[d]{Rogerio Rosenfeld}


\affiliation[b]{
Syracuse University, Physics Department, 
				201 Physics Building, Syracuse, New York, 13244}

\affiliation[c]{
Department of Particle Physics and Astrophysics, Weizmann Institute of Science,
				Herzl St 234, Rehovot 761001, Israel}
\affiliation[d]{
ICTP South American Institute for Fundamental Research and 
Instituto de Fisica Teorica,
UNESP - Universidade Estadual Paulista,
Rua Dr. Bento T. Ferraz 271  -  01140-070  Sao Paulo, SP, Brazil
}
\begin{abstract}
Dilatons are pseudo-Nambu-Goldstone bosons arising from the breaking of conformal invariance.
In this letter we point out that in general a dilaton mass  
has a power-law dependence on a small parameter
related to the explicit breaking of conformal invariance whereas the ratio between the ultraviolet and infrared scales in the theory are exponentially related to the same parameter. We show that this scaling results in a separation between the dilaton mass and the infrared scale that can not be arbitrary large. Therefore a small dilaton mass necessarily is associated to a secluded conformal sector. We argue that the fact that the dilaton field must have a small displacement from the minimum of its effective potential generated near the infrared scale precludes a cosmologically interesting amount of dilatonic dark matter to be produced by a misalignment mechanism in the early Universe. 
\end{abstract}




\end{frontmatter}




\section{Introduction}
\label{introduction}
\label{sec:intro}
Cosmological and astronomical data indicate that there is roughly five times more non-relativistic cold dark matter (CDM) than usual baryonic matter in the universe today.
Although the nature of dark matter is still largely unknown, the most plausible scenario is that it is comprised of new electrically neutral, stable (or metastable with lifetimes longer than the age of the Universe) particles.
The Standard Model (SM) of the electroweak and strong interactions does not have a CDM candidate but several possibilities have been proposed in well-motivated beyond-SM (BSM) scenarios. These possibilities range from ultralight particles with masses well below 1 eV to ultraheavy candidates such as black holes - spanning more than 50 orders of magnitude in the mass of the candidate (see recent reviews on the status of models of sub-eV and super-eV dark matter in \cite{Jaeckel:2022kwg} and \cite{Cooley:2022ufh} respectively).

The prime example of ultralight CDM are axions, pseudo-scalar particles that were first postulated in models constructed to solve the strong CP problem.
Axions or axion-like-particles (ALPs) are pseudo-Nambu-Goldstone bosons of a global $U(1)$ symmetry that is spontaneously broken at a scale $f_a$. The axion mass is protected against large quantum corrections by a remnant shift symmetry.

The relevance of axions and/or ALPs as DM candidates was rapidly recognized as they can lead to the correct relic abundance through the so-called misalignment mechanism where the initial value of the axion field is displaced from its minimum \citep{Preskill:1982cy,Abbott:1982af,Dine:1982ah}. 
For a review of some recent work on the cosmological implications of axions and ALPs as DM in addition to the current and future experimental searches see, {\it e.g.} \cite{Adams:2022pbo}.  Several experiments have been devised and conducted to search for ALPs,
and these searches have already probed a significant portion of the parameter space.  Future experiments using quantum sensors are very promising in further exploring this space. 

An additional interesting possibility for an ultralight spin-0 DM candidate is that of a scalar field. The phenomenology associated with this class of DM models is different from that of the axion case as it is subject to bounds from equivalence principle tests, fifth force searches and leads to an oscillation of quantized energy levels. The sensitivity of experiments to these observables is significantly stronger than those associated with axion DM, which render them attractive from a phenomenological perspective. In fact, in a large fraction of these models the corresponding bounds on the coupling probe super-Planckian values leading to a quality problem~\citep{Banerjee:2022sqg}, see also discussion below. In addition, at higher order in the DM field, even axion-models can be searched for via their quadratic scalar-like coupling~\citep{Kim:2022ype}, although these are naturally more suppressed.

This motivates us to consider models with ultralight scalar DM.
We note, however, that ultralight scalar fields are challenged by naturalness considerations -- quantum corrections typically destabilize the mass of scalar particles.
In the literature three main classes of natural ultralight scalar DM have been considered. One is inspired by the relaxion mechanism~\citep{Graham:2015cka}, where the DM candidate is an axion field with spontaneous CP violation such that it acquires both scalar and pseudo-scalar couplings~\citep{Flacke:2016szy,Banerjee:2018xmn} (see~\cite{Piazza:2010ye} for Higgs-portal models which also share similar features).  
An additional known possibility to protect a light scalar mass naturally is supersymmetry, which typically also predicts 
other light particles (due to the generic presence of complex moduli).

Finally, models with spontaneous breaking of an exact conformal symmetry generates a single massless Nambu-Goldstone boson associated to the dilatation generator and is
referred to as the dilaton (see e.g. \cite{Low:2001bw} and references therein). Explicit breaking of conformal symmetry generates a small mass for the dilaton, turning it into a pseudo-Nambu-Goldstone boson.
Conformal dynamics arises in a broad class of models~\citep{Komargodski:2015grt}.
One celebrated example, motivated by  the hierarchy problem, is the Randall-Sundrum setup \citep{Randall:1999ee}, which is, upon stabilization~\citep{Goldberger:1999uk}, considered to be the holographic dual of an approximately-spontaneously broken conformal field theory~\citep{Arkani-Hamed:2000ijo}. This setup was studied extensively, including the presence of a radion with a mass well below the electroweak scale which is being identified as the holographic dual of the dilaton~\citep{Goldberger:2007zk,Garriga:2002vf,Csaki:2007ns,Csaki:2000zn}. However, in these cases the dilaton is still quite massive and short lived and cannot form the dark matter.  Our focus is to explore sub-eV dark matter, where the electroweak scale is not associated with the conformal sector.

Dilatons indeed have been considered as possible candidates for ultralight dark matter~\citep{Arvanitaki:2014faa}, but mostly from a phenomenological perspective. 
Here we are interested in the more theoretical questions of whether ultralight dilatons can naturally arise in QFT-based models, and whether they can be viable ultralight DM candidates\footnote{The name dilaton has also been used in the context of scalar-tensor theories (eg \cite{Damour:1990tw}) and string theory (eg \cite{Taylor:1988nw}). Here we are referring to a {\it bona fide} dilaton arising from the spontaneous breaking of conformal symmetry.}. 

The possibility of a naturally light dilaton was discussed in the literature. \citep{Grinstein:2011dq,Chacko:2012sy,Bellazzini:2013fga,Coradeschi:2013gda,Megias:2014iwa,Abu-Ajamieh:2017khi}, 
Below we follow the theoretical framework that was established in~\cite{Coradeschi:2013gda}, where it was shown that to obtain a naturally light dilaton  requires the
conformal symmetry to be explicitly broken by near-marginal operators with dimension ($4+\epsilon$), and with the small parameter, $\epsilon$, being related to the dilaton mass. 
We will show that an ultralight radion requires a warp factor much larger than the one needed to solve the hierarchy
between the Planck and weak scales. 
Therefore, if we are interested in an ultralight radion the conformal sector must be unrelated to the SM (see~\cite{Redi:2020ffc,Hong:2022gzo} for discussion of the cosmology of dark conformal sectors). 
In this Letter we point out that an ultralight radion/dilaton arising from a secluded conformal sector
can not be produced at cosmologically relevant abundances via a misalignment mechanism described by a trustworthy effective theory
unless a significant quality issue related to exponentially large transplanckian scales can be circumvented.

\section{Ultralight Dilaton Dark Matter}

We begin our discussion feigning agnosticism about naturalness issues, and focus first only on the most basic phenomenological requirements associated with having coherent dilaton oscillations comprising the majority of cold dark matter.

\begin{center}
    {\bf The Dilaton}
\end{center}
In a secluded conformal sector, in order to have a very light scalar field, we must assume that an approximate conformal symmetry is spontaneously broken.  We associate $f$ with the scale of this breaking, and the field $\phi$ with the corresponding Goldstone boson.

We then parametrize the mass of the dilaton, which arises from some small explicit breaking of conformal symmetry parametrized by $\epsilon$, in terms of the hierarchy between $f$ and this mass:
\begin{equation}
    m_\phi^2 = \epsilon f^2.
\end{equation}

The effective theory of this pseudo-Goldstone boson is valid only to energies as high as $\Lambda \approx 4 \pi f$, beyond which we would require the (here unspecified) dynamics of the unbroken CFT (or the full 5D theory, in the case of a Randall-Sundrum dual description).

\begin{center}
    {\bf  Decoupling the Dilaton}
\end{center}

We first revisit the axion, where the coupling to SM fields is via the divergence of the global PQ current
\begin{equation}
    {\mathcal L}_{\rm a} = \frac{a}{f} \partial_\mu J_\mathrm{PQ}^\mu
\end{equation}
The axion is typically decoupled by simply taking a very large $f$.  Due to the PQ anomaly, the divergence does not vanish, and QCD effects generate axion mass of order $\Lambda_{\rm QCD}^4/f^4$.

Dilaton interactions with other fields are similar, and occur through the coupling to the divergence of the conformal current $J_D^\mu$, and are of the form
\begin{equation}
    {\mathcal L}_{\rm \phi} = \frac{\phi}{f} \partial_\mu J_D^\mu = \frac{\phi}{f} T_{\rm CFT},
\end{equation}
where $T_{\rm CFT}$ is the trace of the stress-energy tensor in the CFT sector.

We shall, however, be interested in $f\ll$ TeV, which implies that the dilaton field needs to be sequesteredby another method.

Sequestration  is required to avoid bounds on long range forces, astrophysics, and collider experiments, and phenomenological viability needs the effective scale suppressing couplings to be greater than the Planck scale~\cite{}.

Fortunately, another mechanism presents itself when the dilaton is part of a strongly coupled dark conformal sector~(\cite{Schmaltz:2006qs}).  In this scenario, the dilaton is a composite state of some rich CFT dynamics.  The SM (+ gravity), which we take to be purely fundamental, is perturbatively coupled to the strong dark sector.  

So at some UV scale $\Lambda$, which we take to be $M_{\rm Pl}$, our theory is expressed schematically as
\begin{equation}
    {\mathcal L} = {\mathcal L}_{\rm SM}+ {\mathcal L}_{\rm CFT} + \sum_i g_i(M_{\rm Pl}) {\mathcal O }_{\rm SM} {\mathcal O}_{\rm CFT}.
\end{equation}
Minimally, the last term \emph{must} include the universal coupling of the fundamental graviton to the stress-energy tensor of the CFT,
\begin{equation}
    \frac{h_{\mu\nu}}{M_{\rm Pl}} T^{\mu\nu}_{\rm CFT} \ni \frac{1}{4} \frac{h}{M_{\rm Pl}} \partial_\mu J^\mu_{D}.
\end{equation}
The trace mode, $h$, is typically pure gauge, however this interaction induces a kinetic mixing of the dilaton in the CFT sector with this trace-mode via exchange of the conformal current, with a mixing angle given approximately by $f/M_{\rm Pl}$.  Consequently, the physical dilaton eigenstate inherits a coupling to the SM fields through this mixing.  Overall, this amounts to an effective additional form factor suppression of $(f/M_{\rm Pl})^2$ in the parts of $\partial_\mu J^\mu_D$ that contain operators of the SM that break conformal symmetry explicitly.  

In the 5D RS dual, this picture is very simple.  Declaring the SM to be purely fundamental amounts to localizing the SM fields on the UV brane, in which case the form factor suppression is simply the wave-function overlap of the radion mode with the UV brane fields.

The coupling between the diagonalized dilaton eigenstate and the fundamental SM fields is therefore given by:
\begin{equation}
{\mathcal L}_d = \left( \frac{f}{M_\mathrm{Pl}} \right)^2 \frac{\phi}{f} T_{\rm SM},
\end{equation}
and the effective scale suppressing the interactions, $M_\mathrm{Pl}^2/f$, is enormously large.  The dilaton is effectively stable, extremely dark, and any long range forces between SM matter are beyond-ultra-weak, suppressed by $(f/M_\mathrm{Pl})$ compared to the gravitational strength.

The confined conformal sector will also have a tower of massive spin-2 excitations beginning near the scale $f$ that mix with the graviton, providing another portal between the SM and the dilaton.  These are also suppressed, and the effects appear at dimension-8 after integrating out this tower of modes:
\begin{equation}
    \frac{1}{M_{\rm Pl}^2 f^2} T^{\mu\nu}_{\rm SM} \partial_\mu \phi \partial_\nu \phi.
\end{equation}
In 5D, these are nothing more than the contributions from KK-graviton exchange, suppressed by the wavefunction overlap with the UV brane.

\begin{center}
    {\bf The Relic Density and the Dilaton Mass}
\end{center}

Since our focus is on scenarios where the dilaton is wave-like, with $m_\phi \lesssim 1$ eV, the simplest mechanism to produce a dark matter relic density in line with observations is that of misalignment.  

The dilaton is therefore taken to have an initial uniform displacement from the minimum of its potential at some early time, and the DM would then be associated with coherent oscillations of the dilaton at frequency $m_\phi$, leading to a present-day density of dark matter given by:
\begin{equation}
    \rho_{\rm DM}=\frac{1}{2} m_\phi^2 \phi_i^2 \left( \frac{a_i}{a_0} \right)^3,
\end{equation}
where $\phi_i$ is the maximium initial displacement of the dilaton field from the minimum of its potential at a time when it is dominantly undergoing free-field harmonic oscillation. The current relic density is obtained from the initial abundance by the usual dilution factor of cold dark matter.

Consistency with cosmological observations and validity of the effective theory of the goldstone boson, $\phi$, puts simple constraints on the hierarchy between $f$ and $m_\phi$.  We now determine these constraints under only two very simple and conservative assumptions:
\begin{itemize}
    \item This DM density must be produced in time to seed cosmological structure formation.  The temperature at which coherent free-field oscillations are established should roughly satisfy $T_i \gtrsim 1$ keV~\citep{Hui:2016ltb}. \footnote{This is likely an overly conservative assumption, since in general scalar fields begin to oscillate at much higher temperatures, $T \sim \sqrt{m_\phi M_p}$. Therefore some mechanism must be invoked to maintain the dilaton field displaced from the minimum of the potential, such as a low-reheat inflation.} 
    \item In order to be in the regime where we have perturbative control over the effective theory while the dilaton is oscillating, we require that at temperature $T_i$, we have $\phi_i \lesssim 4 \pi f$.\footnote{Realistically, this is overly conservative, as the assumption of a quadratic potential is likely a poor approximation that far from the minimum -- self-interactions of the dilaton will be important.  We are interested primarily in an upper bound, however, and this overly conservative assumption thus suits our needs.} 
\end{itemize}

At matter radiation equality, $T_{\rm eq} \approx 3$ eV, we have
\begin{equation}
    \rho_{\rm DM}(T=T_{\rm eq}) =  \frac{1}{2} m_\phi^2 \phi_i^2 \left(T_{\rm eq}/T_i\right)^3\approx 1.15 T_{\rm eq}^4\,\label{rDM}
\end{equation}
and if we substitute $\phi_i \lesssim 4 \pi f= \frac{4 \pi m_\phi}{\sqrt{\epsilon}}$, 
we obtain the following constraint:
\begin{equation}
    \epsilon \lesssim 10^{-8} \left( \frac{m_\phi}{{\rm eV}} \right)^4 \left( \frac{1 {\rm keV}}{T_i} \right)^3.
\end{equation}
Putting this together with the assumption that the dilaton is wave-like and that the initial oscillations begin at $T_i > 1$ keV~\citep{Hui:2016ltb}, we have
\begin{equation}
    \frac{m_\phi}{f} \lesssim 10^{-4}.
\end{equation}

While this seems like a fairly innocent requirement, we will see that the dilaton experiences a quality problem that is exponential in this hierarchy, and that it is difficult to realize $m_\phi/f \lesssim 0.1 - 0.01$.

\section{The quality problem of the ultralight dilaton}

The dilaton in these models suffers from a quality problem that parallels but exponentially exceeds the quality problem of its pseudo-scalar axionic cousin, and that prevents most theories from naturally realizing ultra-light scalar fields.

For the axion, the quality problem is related to the folk theorem that quantum gravity effects violate all global symmetries.  For the effective field theorist, this corresponds to the necessity for inclusion of higher-dimensional operators, suppressed by appropriate powers of the Planck scale and with $\mathcal{O}(1)$ Wilson coefficients, that contribute to an explicit breaking of the PQ symmetry that is distinct from the PQ anomaly.  Such operators contribute to the axion potential, and generically displace the axion vacuum expectation value from the CP-conserving state when they are comparable to the size of instanton effects.  

For example, at dimension-5, with coupling constant $g$, we would require 
\begin{equation}
    \frac{g f_a^5}{\Lambda_\mathrm{UV}} \lesssim \Lambda_\mathrm{QCD}^4 \times \epsilon_{\rm edm} \mathrm{~~~~~or~~~~~} \Lambda_\mathrm{UV} \gtrsim \frac{g f_a^5}{\Lambda_\mathrm{QCD}^4 \times \epsilon_{\rm edm}}
\end{equation}
in order for the axion solution to the QCD CP-problem (which require $\epsilon_{\rm edm}\lesssim 10^{-10}$) to remain unspoiled.
For an axion decay constant $f_a \sim 10^9$ GeV, this corresponds to $\Lambda_\mathrm{UV} \gtrsim 10^{57}$ GeV for $g\sim\mathcal{O}(1)$, well above the Planck scale.  Solutions to the axion quality problem must suppress the Wilson coefficients of these operators to some high dimensionality through symmetry or dynamics.

The difficulties associated with obtaining an ultralight dilaton can be made clear from studying the low energy effective theory of spontaneously broken conformal invariance.  In a model with no explicit breaking of conformal symmetry, the dilaton EFT contains an invariant quartic coupling which destabilizes any time-independent background solution unless the quartic is vanishing.  Thus explicit breaking of conformal invariance must be introduced in order to obtain a unique classical vacuum -- the dilaton is down-graded to an approximate goldstone boson.

The potential for the vacuum expectation value of the dilaton can therefore be schematically represented as
\begin{equation}
V(\chi) = \lambda \chi^4 \rightarrow \lambda(\chi) \chi^4
\end{equation}
where $\chi$ is the canonically normalized dilaton field, and on the right hand side, explicit breaking of conformal invariance is introduced through a $\chi$ dependence in the conformally invariant quartic.

With the definition $\beta \equiv \frac{d\lambda}{d\log\chi}$, extremization of this potential occurs when $\beta_* = -4\lambda_*$, where the starred quantities are the $\beta$-function and coupling evaluated at the value of the dilaton decay constant, $\chi_*  = \langle \chi \rangle \equiv f$.

The physical dilaton corresponds to fluctuations of $\chi$ about its VEV, $f$:  $\chi(x) = f + \phi(x)$.  Its mass is then obtained from taking the second derivative of the potential and imposing the minimization condition, $\lambda_* = -\frac{1}{4} \beta_*$:
\begin{equation}
    m^2_\phi = \left[ \left(\frac{ d \log \beta}{d \log \chi}\right)_* + 4 \right] \beta_* f^2.
\end{equation}

There is a clear tension in having a large suppression in the dilaton mass relative to the scale $f$.  For the dilaton to be ultralight, the $\beta$-function must be extremely small at the scale $f$.  The extremization condition, however, links the $\beta$-function to the value of the quartic, which must therefore also be tiny.  

There are thus two options:  tune the quartic to be relatively small, which is the case in Goldberger-Wise type stabilization scenarios, or allow there to be a large $\beta$-function associated with a coupling that is very small in the UV, which is explored in \cite{Csaki:2023pwy}.

An ultralight dilaton therefore requires a \emph{conspiracy} that both $\beta$ and $\lambda$ be very small at the scale $f$.  It is not especially difficult to realize a small $\beta$ function at, e.g, an IR fixed point, but a small $\lambda$ corresponds to a near flat-direction in a scalar potential.  Without additional symmetries (such as supersymmetry), there is no reason that this IR fixed point should coincide with an extremely flat scalar potential.

An idea to circumvent this was to have a very large range of $\chi$ over which the $\beta$-function remains very small.  In this case, the coupling $\lambda$ slowly runs over a large range of scales, and eventually becomes small, triggering the breaking of conformal symmetry when $\beta = -4 \lambda$.  This would remove the fine-tuning issue of requiring a small quartic coincident with slow running.

The quality problem with this proposal is manifest in the solution to the corresponding renormalization group equation.  Parameterizing the $\beta$-function as $\beta = \epsilon g(\chi)$, where $\epsilon$ is a small constant, and g is an order one function of $\chi$, the solution is
\begin{equation}
   \Lambda_\mathrm{UV} \approx f \exp\left[ \left|\frac{\lambda(\Lambda_\mathrm{UV}) - \lambda_*}{\epsilon}  \right| \right].
\end{equation}
If $\epsilon$ is taken to be very small, the initial UV scale in the running must be deeply trans-planckian.

However, the largest we might consider taking $\Lambda_\mathrm{UV}$ to be is the Planck scale, where gravitational effects explicitly violate conformal symmetry, akin to how they are expected to break global symmetries.  Above this scale, higher dimensional operators that explicitly break conformal symmetry become important, and the assumption of a small $\beta$-function is no longer tenable without a full UV description.  

To leading Log we find that 
$m_\phi^2/f^2\approx\lambda(\Lambda_{\rm UV})/\ln(\Lambda_{\rm UV}/m_\phi)\,.$
If we take the dilaton mass to be sub-eV, and the UV value of the quartic to be ${\mathcal O} (1-4\pi)$, then we find that the smallest we can plausibly take $\epsilon$ to be is
\begin{equation}
    \frac{m^2_\phi}{f^2} \sim 0.01 - 0.1.
\end{equation}
Due to the exponential sensitivity to $\epsilon$, this constraint applies relatively consistently across the entire range of dilaton DM masses that we consider, $10^{-22} - 1$ eV.

The same conclusions are obtained in a dual description of spontaneously broken conformal symmetry, the 5-dimensional Randall-Sundrum I model with UV and IR branes. The warped geometry is responsible for the hierarchy between the energy scales in the UV and IR branes and the Planck/weak scale hierarchy
is obtained with a mild fine tuning of the radius of the 5th warped dimension which fixes the separation between the branes.
However, this radius is undetermined in the model, being in fact a flat direction. The 5D gravitational fluctuations contain a scalar degree of freedom called the radion, which is the holographic dual of the dilaton.
However, it has no potential and corresponds to a massless particle in the original model. In order to determine the radius of the extra dimension, given by the vacuum expectation value of the radion field, a potential for the radion was obtained by Goldberger and Wise with the introduction of a massive scalar field propagating in the bulk.
This scalar field has potentials localized on the UV and IR branes that lead to vacuum expectation values of the 5D scalar on the branes, $v_{UV}$ and $v_{IR}$.
The ratio $\varepsilon \equiv m^2/4 k^2$  between the mass of the scalar field $m$ and the inverse of the AdS curvature $k$, can be a small parameter, and an effective potential for the radion is then generated through gravitational backreaction of the scalar field on the background metric:
\begin{equation}
    V_{eff}(r) = 4 k e^{-4 \pi k r} \left( v_{IR} - v_{UV} e^{- \varepsilon \pi k r}    \right)^2 + {\cal O}(\varepsilon^2)
\end{equation}
which has a nontrivial minimum at: 
\begin{equation}
r = \frac{1}{\varepsilon \pi k} \ln\left( \frac{v_{UV}}{v_{IR}} \right).
\end{equation}
This leads to a radius stabilization and a hierarchy:
\begin{equation}
\frac{\Lambda_{UV}}{\Lambda_{IR}} =  e^{ \pi k r}   = \left(  \frac{v_{UV}}{v_{IR}} \right)^{1/\varepsilon}.    
\end{equation}
The mass of the dilaton has to be determined from a correct identification of the canonically normalized radion field 
($\chi (x) \sim e^{- \pi  k r(x)} $) and a full computation of a 5-dimensional scalar-gravity system, taking into account the backreaction of the radion perturbations on the AdS metric. To leading order in the backreaction the dilaton mass can be written schematically  as \citep{Kofman:2004tk}
\begin{equation}
    m_d^2 \sim \varepsilon \Lambda_{IR}^2,
\end{equation}
In summary, the fact the the dilaton mass has a power-like dependence whereas the hierarchy has an exponential dependence on the small parameter characterizing the breaking of scale symmetry seems to be quite general.
As a consequence, it is not possible to obtain an arbitrarily large separation ($\varepsilon \ll 1$) between the dilaton mass and the symmetry beaking scale.
This scaling  precludes an ultralight dilaton unless a very large hierarchy between the ultraviolet and infrared scales can be tolerated.

\section{Possible generalizations}
One might question whether the quality problem we have observed is an artifact of focusing on too simple a model, where we have envisioned that a single operator leads to stabilization.~\footnote{We thank Zohar Komargodski for suggesting that we explore this possibility.}  Are there (non-supersymmetric) generalizations that escape the tuning for a light dilaton?  We argue here that this is not the case.

The essential reason behind the tuning is that, in a non-supersymmetric theory, a non-compact scalar flat direction is unprotected.    A light dilaton requires that, at the scale $f$, there is a conspiracy of small explicit breaking of conformal symmetry and a scalar flat direction.  This is the case regardless of how many operators are contributing to the effective running.

More concretely, we might consider holographic models with multiple scalar bulk fields contributing to stabilization \citep{Eroncel:2019zev}. For instance, for two different small deformations characterized by the parameters $\varepsilon_1$ and $\varepsilon_2$ one obtains an effective potential of the form:
\begin{equation}
    V(\phi) =  \phi^4 \left[ \lambda + \lambda_1 \left( \frac{\phi}{\langle \phi \rangle} \right)^{\varepsilon_1} + \lambda_2 \left( \frac{\phi}{\langle \phi \rangle} \right)^{\varepsilon_2}   \right],
\end{equation}
The minimization condition can be used to express $\lambda$ in terms of $f$, in which case the dilaton mass has a simple form:
\begin{equation}
    m_\phi^2 \approx 4 f^2 \left[ \varepsilon_1 \lambda_1 f^{\varepsilon_1}+ \varepsilon_2 \lambda_2 f^{\varepsilon_2} \right]
\end{equation}
To obtain a dilaton mass that is parametrically small compared to either of the $\beta$-functions thus requires a cancelation between these two terms.  It is in principle possible to fine-tune these extra parameters in order to obtain a large scale separation between the dilaton mass and the IR scale. However, we consider such a tuning to be unnatural.

\section{Summary and Conclusions}
In this Letter we point out obstacles associated with obtaining natural ultra-light scalar fields as the dominant contribution to the DM density of our Universe.  The most promising theoretical candidate for ultra-light scalars is the pseudo-goldstone boson associated with the spontaneous breaking of conformal symmetry, the dilaton, and so our discussion has focused on these models.

We first noted that sequestration of the dilaton is not an issue, due to an enormous form-factor suppression of the composite dilaton interactions with SM fields.  This arises due to simple wave-function overlap suppression in the 5D dual picture.

In addition, we also show that the requirement of a small initial dilaton displacement necessary for the validity of the
low-energy effective description of the theory constrains the ratio $(m_\phi/f)^2 \lesssim 10^{-8}$.

However, theories describing ultralight dilatons with a small explicit breaking of scale invariance parametrized by $\varepsilon \sim (m_\phi/f)^2$ 
present a power-law scaling for the dilaton mass and a exponential-law scaling for the ratio between the UV and IR energy scales
in this parameter.  
We show that it is not possible to obtain a large gap between the dilaton mass and the dilaton decay constant without introducing
a severe quality problem reflected in a huge hierarchy between the infrared and ultraviolet scales.  To avoid this problem, we need $\varepsilon \lesssim 0.1 - 0.01$.  We thus conclude that there is a severe conflict between light dilaton naturalness and phenomenological constraints on the misalignment hypothesis for dark matter production.

\section*{Acknowledgements}
The authors are grateful to D.~E.~Kaplan for many discussions and contributions when the core of the work related to this work was made.
The authors also thank A.~Banerjee for discussions during the initial phase of the paper, and comments on the manuscript, and we thank Z.~Komargodski and R.~Rattazzi for discussions.
The work of JH is supported in part by U.S. Department of Energy (DOE), Office of Science, Office of High Energy Physics, under Award Number DE-SC0009998.  The work of GP is supported by grants from BSF-NSF, Friedrich Wilhelm Bessel research award of the Alexander von Humboldt Foundation, GIF, ISF, Minerva,
SABRA - Yeda-Sela - WRC Program, the Estate of Emile Mimran, and the Maurice and Vivienne Wohl Endowment.
The work of RR is supported in part by the São Paulo Research Foundation (FAPESP) through grant \# 2021/10290-2 and by the Brazilian Agency CNPq through a productivity grant \# 311627/2021-8. RR thanks the hospitality of the Department of Particle Physics and Astrophysics at the Weizmann Institute of Science, where this work was initiated with the support of an Erna and Jakob Michael Visiting Professorship.

\bibliographystyle{elsarticle-harv} 
\bibliography{example}






\end{document}